\documentstyle[times]{mn}
\newread\epsffilein    
\newif\ifepsffileok    
\newif\ifepsfbbfound   
\newif\ifepsfverbose   
\newdimen\epsfxsize    
\newdimen\epsfysize    
\newdimen\epsftsize    
\newdimen\epsfrsize    
\newdimen\epsftmp      
\newdimen\pspoints     
\def\epsfbox#1{\global\def\epsfllx{72}\global\def\epsflly{72}%
   \global\def\epsfurx{540}\global\def\epsfury{720}%
   \def\lbracket{[}\def\testit{#1}\ifx\testit\lbracket
   \let\next=\epsfgetlitbb\else\let\next=\epsfnormal\fi\next{#1}}%
\def\epsfgetlitbb#1#2 #3 #4 #5]#6{\epsfgrab #2 #3 #4 #5 .\\%
   \epsfsetgraph{#6}}%
\def\epsfnormal#1{\epsfgetbb{#1}\epsfsetgraph{#1}}%
\def\epsfgetbb#1{%
\openin\epsffilein=#1
\ifeof\epsffilein\errmessage{I couldn't open #1, will ignore it}\else
   {\epsffileoktrue \chardef\other=12
    \def\do##1{\catcode`##1=\other}\dospecials \catcode`\ =10
    \loop
       \read\epsffilein to \epsffileline
       \ifeof\epsffilein\epsffileokfalse\else
          \expandafter\epsfaux\epsffileline:. \\%
       \fi
   \ifepsffileok\repeat
   \ifepsfbbfound\else
    \ifepsfverbose\message{No bounding box comment in #1; using defaults}\fi\fi
   }\closein\epsffilein\fi}%
\def\epsfsetgraph#1{%
   \epsfrsize=\epsfury\pspoints
   \advance\epsfrsize by-\epsflly\pspoints
   \epsftsize=\epsfurx\pspoints
   \advance\epsftsize by-\epsfllx\pspoints
   \epsfxsize\epsfsize\epsftsize\epsfrsize
   \ifnum\epsfxsize=0 \ifnum\epsfysize=0
      \epsfxsize=\epsftsize \epsfysize=\epsfrsize
     \else\epsftmp=\epsftsize \divide\epsftmp\epsfrsize
       \epsfxsize=\epsfysize \multiply\epsfxsize\epsftmp
       \multiply\epsftmp\epsfrsize \advance\epsftsize-\epsftmp
       \epsftmp=\epsfysize
       \loop \advance\epsftsize\epsftsize \divide\epsftmp 2
       \ifnum\epsftmp>0
          \ifnum\epsftsize<\epsfrsize\else
             \advance\epsftsize-\epsfrsize \advance\epsfxsize\epsftmp \fi
       \repeat
     \fi
   \else\epsftmp=\epsfrsize \divide\epsftmp\epsftsize
     \epsfysize=\epsfxsize \multiply\epsfysize\epsftmp   
     \multiply\epsftmp\epsftsize \advance\epsfrsize-\epsftmp
     \epsftmp=\epsfxsize
     \loop \advance\epsfrsize\epsfrsize \divide\epsftmp 2
     \ifnum\epsftmp>0
        \ifnum\epsfrsize<\epsftsize\else
           \advance\epsfrsize-\epsftsize \advance\epsfysize\epsftmp \fi
     \repeat     
   \fi
   \ifepsfverbose\message{#1: width=\the\epsfxsize, height=\the\epsfysize}\fi
   \epsftmp=10\epsfxsize \divide\epsftmp\pspoints
   \newcount\figskipcount
      \message{#1 \the\epsfysize  }
   \vbox to\epsfysize{\vfil\hbox to\epsfxsize{%
      \includegraphics{#1}%
      \hfil}}%
\epsfxsize=0pt\epsfysize=0pt}%

{\catcode`\%=12 \global\let\epsfpercent=
\long\def\epsfaux#1#2:#3\\{\ifx#1\epsfpercent
   \def\testit{#2}\ifx\testit\epsfbblit
      \epsfgrab #3 . . . \\%
      \epsffileokfalse
      \global\epsfbbfoundtrue
   \fi\else\ifx#1\par\else\epsffileokfalse\fi\fi}%
\def\epsfgrab #1 #2 #3 #4 #5\\{%
   \global\def\epsfllx{#1}\ifx\epsfllx\empty
      \epsfgrab #2 #3 #4 #5 .\\\else
   \global\def\epsflly{#2}%
   \global\def\epsfurx{#3}\global\def\epsfury{#4}\fi}%
\def\epsfsize#1#2{\epsfxsize}
\def\figinsert#1#2{\epsfbox{#1} \message{#2} }    
\begin{document}

\def\propsima{$\; \buildrel \sim \over \propto   \;$}
\def\propsim{\lower.5ex\hbox{\propsima}}

\title
[X-ray variability in a deep, flux limited sample of QSOs]
{X-ray variability in a deep, flux limited sample of QSOs}
\author[O. Almaini et al. ]
{O.~Almaini,$^{1}$
A. Lawrence,$^1$ 
T.~Shanks,$^2$ 
A. Edge,$^2$ 
B.J.~Boyle,$^3$ 
I.~Georgantopoulos,$^4$  \cr
K.F. Gunn,$^5$
G.C. Stewart,$^6$
and R.E.~Griffiths$^7$ 
\\
$^1$ Institute for Astronomy, Royal Observatory, 
University of Edinburgh, Blackford
Hill, Edinburgh EH9 3HJ \\
$^2$ Department of Physics, University of Durham, South Road,
Durham DH1 3LE  \\
$^3$ Anglo-Australian Observatory, PO Box 296, Epping NSW 2121, Australia \\
$^4$ National Observatory of Athens, I. Metaxa \& B. Pavlou, Palaia Penteli, 
15236, Athens, Greece \\
$^5$ Department of Physics \& Astronomy, University of Southampton,
University Road, Southampton SO17 1BJ \\
$^6$ Department of Physics and Astronomy, The University of Leicester, 
Leicester LE1 7RH \\
$^7$ Department of Physics, Carnegie Mellon University, Wean Hall, 
5000 Forbes Ave., Pittsburgh, PA 15213, USA }
\date{MNRAS in press}
\maketitle

\begin{abstract}
We present an analysis of X-ray variability in a flux limited sample
of QSOs. Selected from our deep ROSAT survey, these QSOs span a wide
range in redshift ($0.1<z<3.2$) and are typically very faint, so we
have developed a method to constrain the amplitude of variability in
ensembles of low signal-to-noise light curves. We find evidence for
trends in this variability amplitude with both redshift and
luminosity. The mean variability amplitude declines sharply with
luminosity, as seen in local AGN, but with some suggestion of an
upturn for the most powerful sources. We find tentative evidence that
this is caused by redshift evolution, since the high redshift QSOs
($z>0.5$) do not show the anti-correlation with luminosity seen in
local AGN. We speculate on the implications of these results for
physical models of AGN and their evolution. Finally, we find evidence
for X-ray variability in an object classified as a narrow
emission-line galaxy, suggesting the presence of an AGN.

\end{abstract}

\section{Introduction}

The X-ray flux from AGN exhibits variability on shorter timescales
than any other waveband, indicating that this emission occurs in the
innermost region of the central engine. Detailed studies of X-ray
variability can therefore act as important probes of the AGN
phenomenon, potentially yielding information on size scales
unresolvable by any other means. 

While previous observations had revealed that X-ray variability was
very common in AGN (see Mushotzky, Done \& Pounds 1993 for a review),
the most significant advances were made with the `long look'
observations taken with the $\em EXOSAT$ observatory. The highly
eccentric orbit of this satellite combined with its much improved
sensitivity allowed continuous observations of a small number of local
AGN with high signal to noise.  The light curves appeared very random
in nature.  Fourier and spectral analyses revealed no preferred
timescales or periodicities. Instead these sources showed power
smoothly distributed over a wide range of frequencies, generally with
power spectra of the form $P(f) \propto f^{-1.5}$ (Lawrence \&
Papadakis 1993, Green et al 1993). Such light curves are described as
$\em red\,noise$, i.e. scale free but with more power at low
frequencies.  The power spectra are eventually expected to flatten
from a power-law form at the lowest frequencies below $\sim 10^{-6}$
Hz (Edelson et al 1999). Similarly, at high
frequencies we expect a cutoff corresponding to the smallest scales in
the emitting region (e.g. the inner region of the accretion disk) but
so far no such steepening has been observed.

Interesting results were obtained by comparing the variability in
different AGN, suggesting that more luminous sources vary with lower
amplitude. Barr \& Mushotzky (1986) studied the shortest timescale
required for a source to double in flux and observed an increase in
the doubling timescale with luminosity. Lawrence \& Papadakis (1993)
and Green et al (1993) studied the power spectra of the $\em EXOSAT$
AGN and found that the power at a given frequency decreases with
luminosity with the form $P \propto L_x^{-{\rm(0.6\pm0.1)}}$.  This
result has since been confirmed by observations with the $\em ASCA$
satellite, albeit with overlapping samples of AGN (Nandra et al 1997,
Turner et al 1999).  Such trends might be explained if more luminous
sources are physically larger in size.  However, these studies were
conducted on a small number of bright AGN at low redshift ($z<0.1$).
Since QSO activity is known to have peaked at much earlier epochs
($z\simeq 2$) it is now vital to establish whether the trends observed
in these local AGN apply to the QSO population as a whole and to test
for evolutionary effects.  In this work we present an analysis of a
flux limited sample of 86 QSOs detected as part of the Deep {\em
ROSAT} Survey (Georgantopoulos et al 1996). Spanning a wide range in
redshift ($0.1<z<3.2$), the same sample has been used to define the
form and evolution of the QSO luminosity function (Boyle et al
1994). Most of the sources are detections at the faintest possible
X-ray flux and hence the light curves are generally of low signal to
noise.  Nevertheless, we have developed methods for constraining the
level of intrinsic variability in each source. By combining the
results from many QSOs we can search for trends in the typical
variability amplitude with redshift and luminosity. Finally we study
the variability of X-ray emitting galaxies to test the hypothesis that
these contain obscured AGN.

\section{The sample}

Our sample consists of 86 QSOs identified during the Deep {\em ROSAT}
Survey of Shanks et al (in preparation), which reached an X-ray flux
limit of $\sim4\times10^{-15}$erg$\,$s$^{-1}$cm$^{-2}$ (0.5-2.0 keV).
A description of the X-ray source detection and optical identification
is given in Georgantopoulos et al (1996). Details of the QSO redshift
distribution and the X-ray luminosity function and its evolution are
given in Boyle et al (1994).

The survey uses 7 deep {\em ROSAT} PSPC pointings, with integration times of
$\sim 30-80$ks per field, spread over periods ranging from 2-14 days.
Periodic gaps in the light curves are caused by the orbit of the
satellite, with larger gaps occurring during phases of high instrument
background, such as when the satellite passes through the South
Atlantic Anomaly.  During each 96 minute orbit, a source is typically
visible for $1000-2000$ seconds. Given the faint nature of the X-ray
sources, light curves were first binned on these periodic orbits,
giving $16-26$ observation slots (depending on the field).  The
fainter sources are then re-binned to allow meaningful Gaussian
statistics.  Using the mean intensity of the source, if the smallest
bin is expected to give fewer than 20 photons it is merged with its
nearest neighbour.  This process is repeated until $\bar{I} \Delta t >
20 $ photons per bin, leaving 86 QSOs with an absolute minimum of at
least 2 time bins.  We note that this re-binning process will
inevitably smear out high frequency variability, and discuss this
effect further in Section 3 and Appendix A.

Basic data reduction was carried out using the $\em Asterix$ X-ray
data processing package.  For each source, depending on the off-axis
radius, a source box radius was chosen to enclose $90$ per cent of the
X-ray photons. Assuming a mean QSO spectral index of $\Gamma=2.2$
(Almaini et al 1996), the radius of this circle varies from $24.6''$
on axis to $56.6''$ at a maximum radius of 18 arcmin.  Data from
periods of high particle background are excluded from the analysis.
This removes $\sim 10$ per cent of the data when the Master Veto Rate
rises above 170 counts s$^{-1}$ (Plucinsky et al 1993). Considerable
care was taken in choosing areas for background subtraction. By good
fortune, none of the fields showed evidence for significant background
gradients (generally caused by irregularities in the galactic
background or contamination from solar scattered X-rays).  Typically
10 circular areas of 4-6 arcmin radius are then chosen across each
field from source free regions to perform the background subtraction.
The background box closest to each source was used, with minor
vignetting corrections ($< 5$ per cent) to correct for the differing
off axis radii of source and background boxes.

Problems may also conceivably arise if the background shows differing
variability across the field. Although care is taken to use background
regions close to each source, differential variability may
nevertheless introduce extra variance into the light curves above the
Poisson noise and bias the results.  Some of the fields do indeed show
evidence for variability at a low level ($\delta I/I < 10$ per cent),
probably due to contamination from solar scattered X-rays, but this is
consistent across the detector.  Such minor effects will therefore be
removed in background subtraction. Finally, to check that any
variability was real and not due to a flickering response in the PSPC, we
compared the light curves of brighter QSOs with each other and with
background regions. No significant correlations could be seen,
suggesting that the bulk of the variability is indeed real and not
caused by mysterious instrumental effects.

\begin{figure}
\caption{Light curve and likelihood function (of arbitrary
normalisation) for the QSO GSGP4X:015} \centering
\centerline{\epsfxsize=5.1 truecm \figinsert{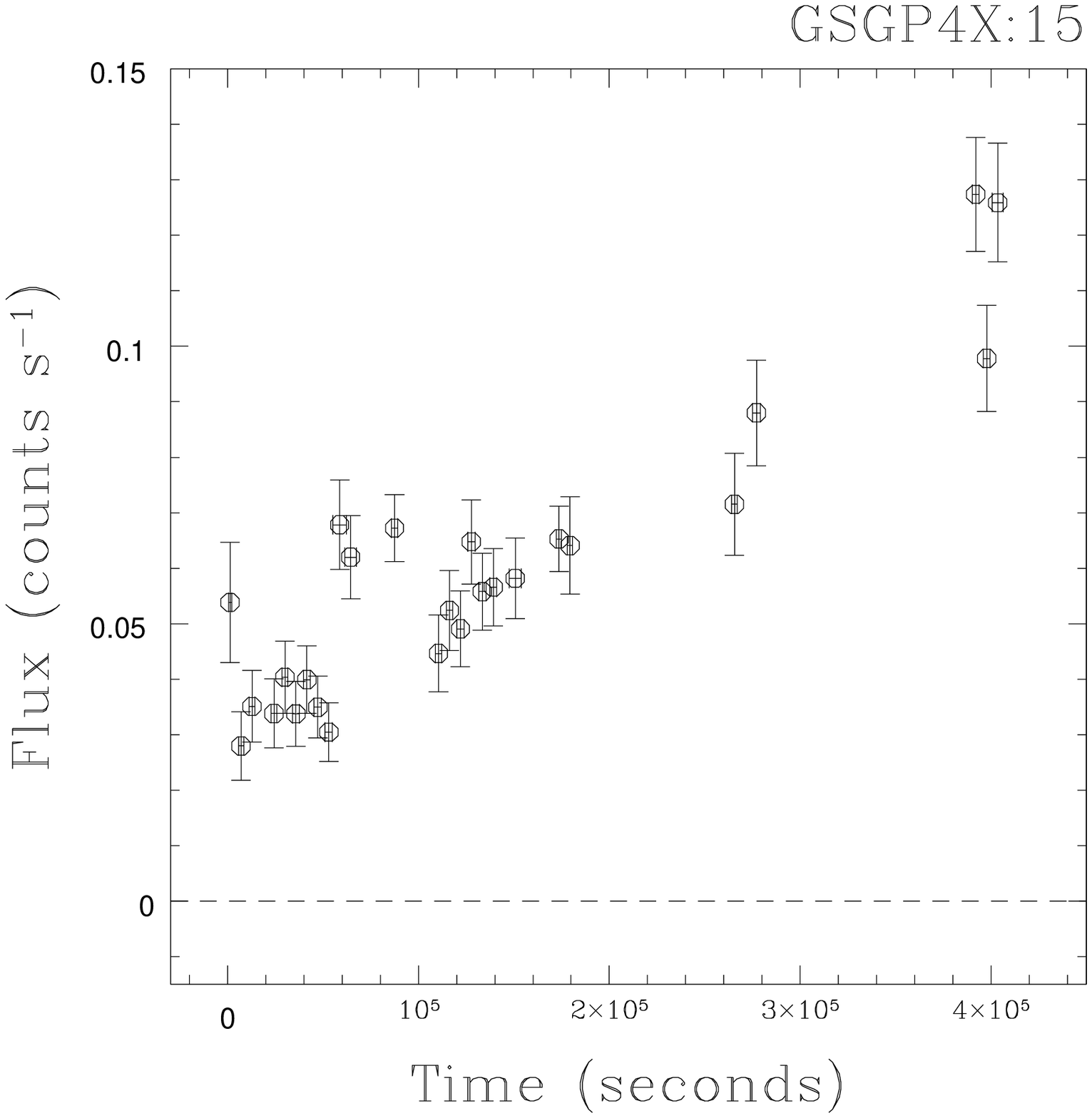}{0.0pt} }
\centerline{\epsfxsize=5.1 truecm \figinsert{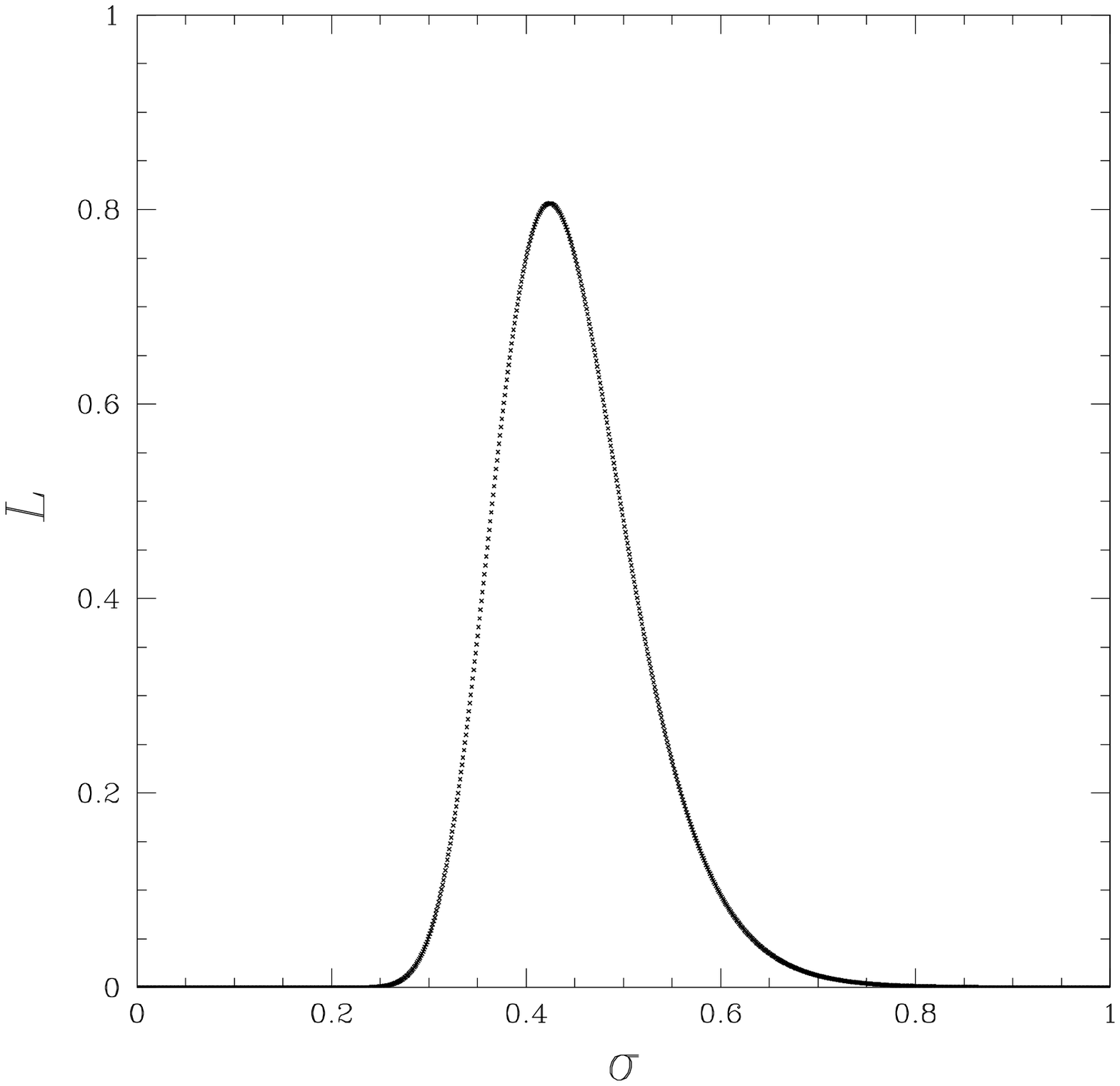}{0.0pt}
}
\end{figure}

\begin{figure}
\caption{Light curve and likelihood function (of arbitrary
normalisation) for the QSO GSGP4X:040} \centering
\centerline{\epsfxsize=5.1 truecm \figinsert{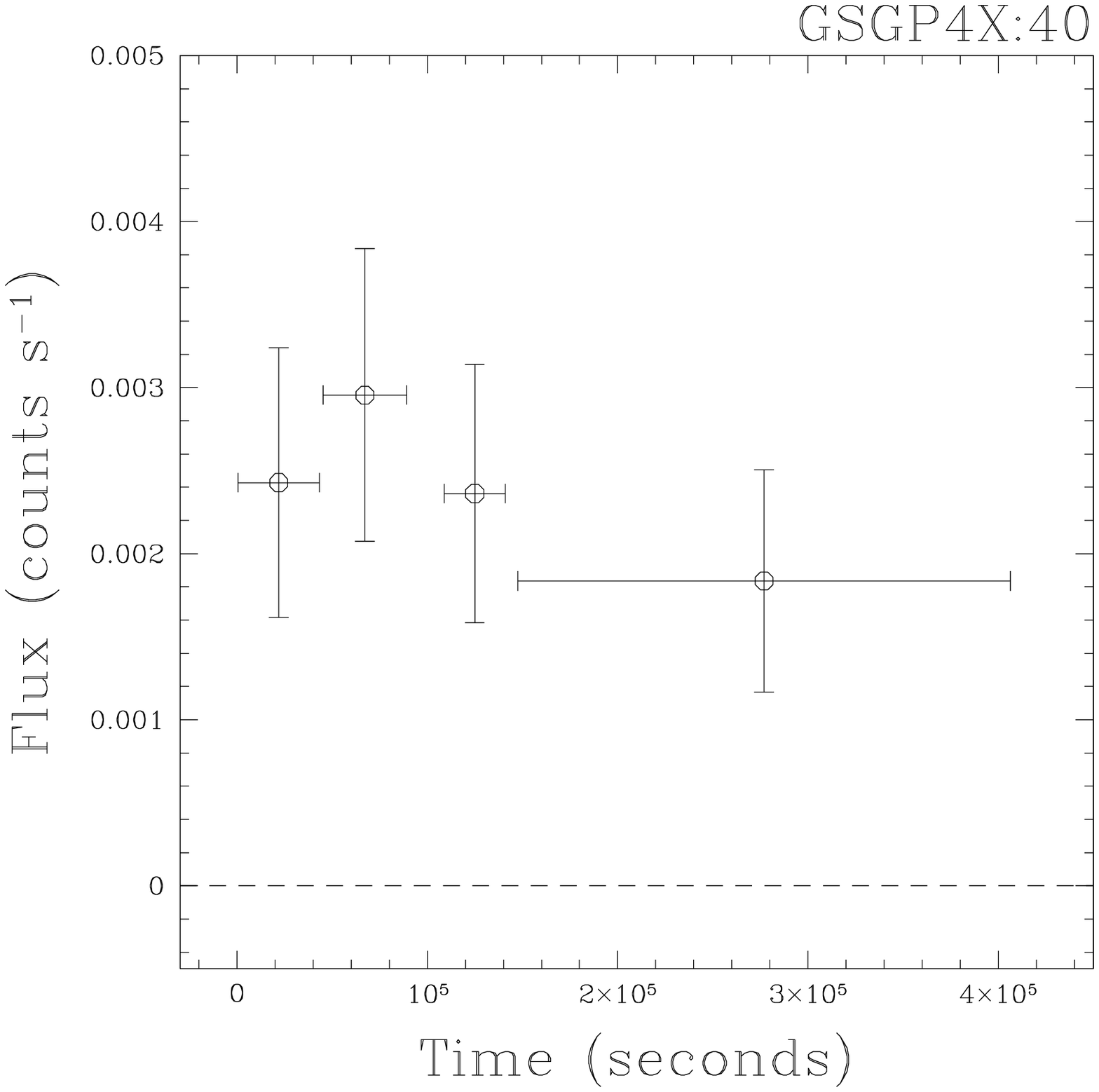}{0.0pt} }
\centerline{\epsfxsize=5.1 truecm \figinsert{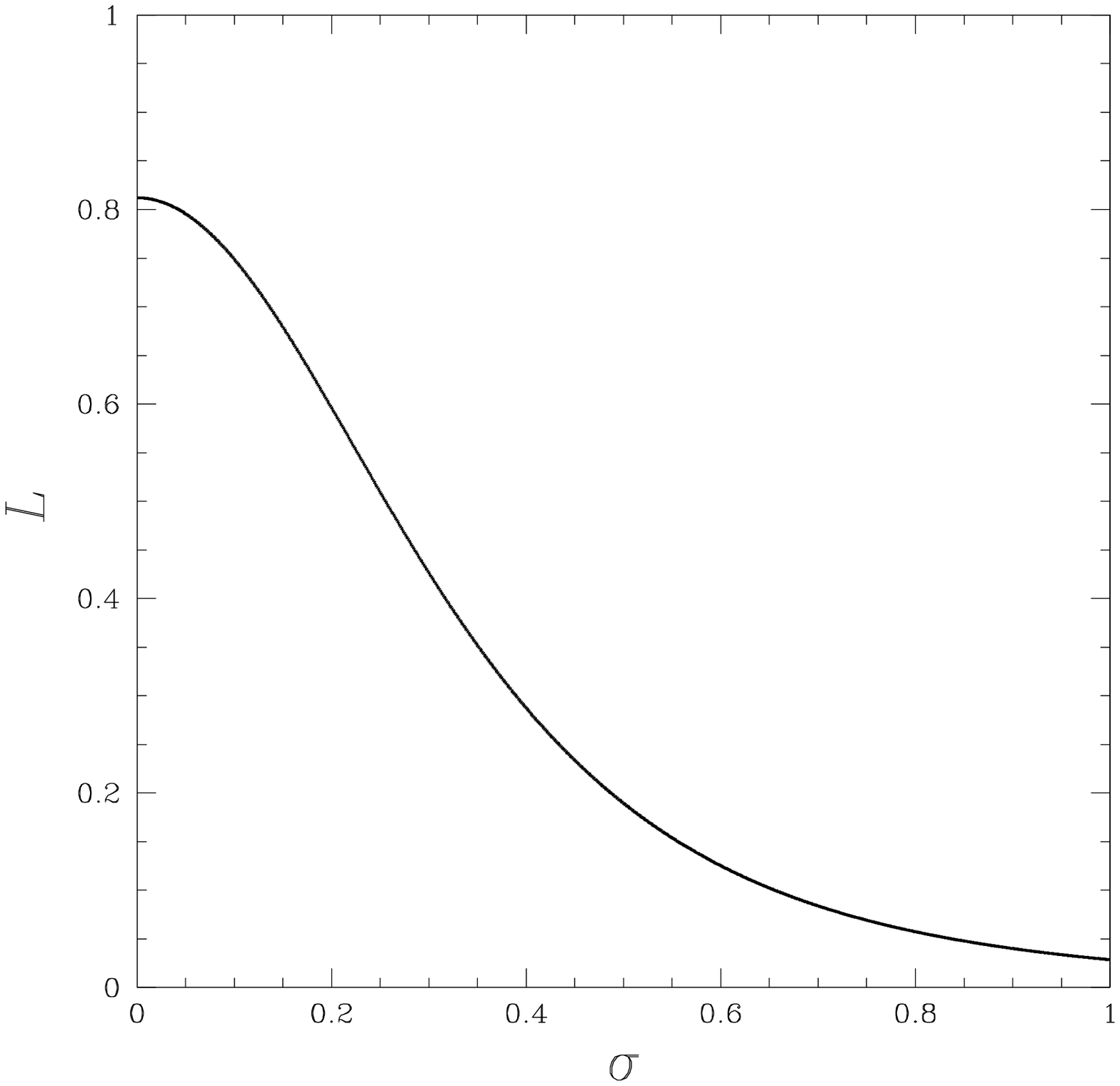}{0.0pt}
}
\end{figure}

\section{Measuring the amplitude of QSO variability}

We now develop a method for measuring the amplitude of intrinsic
fluctuations in light curves of low signal to noise. Although many
will be too faint to enable a significant detection of variability, we
can nevertheless place upper limits on the amplitude of fluctuations
which could be present. In order to combine information from faint
light curves, we will also extend this method to measure the typical
variability in an ensemble of QSOs (e.g. within a given luminosity or
redshift range).

Unlike the well studied $\em EXOSAT$ light curves, the QSOs studied
here are too faint to obtain variability amplitudes by detailed power
spectrum analysis. A simpler method is to estimate the intrinsic
variance in the light curve. Since this quantity is proportional to
the amplitude of the power spectrum (see Appendix A) it will also
allow comparisons with previous work. Nandra et al (1997) used the
root mean square variation in their $\em ASCA$ lightcurves, subtracting off
a contribution due to noise. An intrinsic assumption in this method is
that each point in the light curve has equal weight, which will
break down in cases where the errors differ significantly from point
to point. We therefore develop a new maximum likelihood technique to
extract the best estimate of the intrinsic (noise subtracted) variance
given a set of data values $x_i$ with differing measurement errors
$\sigma_i$.

In order to compare objects of different flux in a meaningful way, all
light curves are first divided by their mean flux.  Hence we are
measuring the amplitude of $\em fractional$ variability. When an
ensemble of QSOs is considered (e.g. all QSOs within a given redshift
range) we treat each light curve as a `snapshot' from the population
and assume that the ensemble all have the same underlying distribution
in $I/\bar{I}$. By definition the variance does not depend on the
ordering of the points and hence we can combine the light curves and
measure the variance in the resulting distribution. We now describe a
method for separating the intrinsic variance from the scatter due to
measurement errors.

\subsection{A maximum likelihood estimator}

The variance in a light curve has two components: first there will be
fluctuations due to noise ($\sigma^2_{noise}$) and secondly there may
be intrinsic variations from the QSO or ensemble of QSOs
($\sigma^2_Q$). Hence:

\begin{equation}
\sigma^2_{total}=\sigma^2_{noise}+\sigma^2_{Q}
\end{equation}

If we assume Gaussian statistics, for a light curve with mean
$\bar{x}$ and measurement errors ${\sigma_i}$, the probability density
for obtaining the N data values ${x_i}$ is given by:

\begin{equation}
p(x_i|\sigma_i,\sigma_Q)=\prod_{i=1}^N\frac{\exp\left\{-\frac{1}{2}
(x_i-\bar{x})^2/(\sigma_i^2+\sigma_Q^2)\right\}}
{(2\pi)^{1/2}(\sigma_i^2+\sigma_Q^2)^{1/2}}
\end{equation}

This is simply a product of $N$ Gaussian functions representing the
probability distribution for each bin. We may turn this around using
Bayes' Theorem to obtain the probability distribution for $\sigma_Q$
given our measurements:

\begin{eqnarray}
p(\sigma_Q|x_i,\sigma_i)& =&
p(x_i|\sigma_i,\sigma_Q)\frac{p(\sigma_Q)}{p(x_i)} \\ & \propto &
L(\sigma_Q|x_i,\sigma_i)
\end{eqnarray}

where $L(\sigma_Q|x_i,\sigma_i)$ is the likelihood function for the
parameter $\sigma_Q$ given the data. This general form for the
likelihood function can be calculated if one assumes a Bayesian prior
distribution for $\sigma_Q$ and $x_i$. In the simplest case of a
uniform prior one obtains:

\begin{eqnarray}
L(\sigma_Q|x_i,\sigma_i) & \propto & p(x_i|\sigma_i,\sigma_Q) \\ & = &
                         \prod_{i=1}^N\frac{\exp\left\{-\frac{1}{2}
                         (x_i-\bar{x})^2/(\sigma_i^2+\sigma_Q^2)\right\}}
                         {(2\pi)^{1/2}(\sigma_i^2+\sigma_Q^2)^{1/2}}
\end{eqnarray}

By differentiating the above, the maximum likelihood estimate for
$\sigma_Q$ can be shown to satisfy the following, which (for a uniform
prior) is mathematically identical to a least $\chi^2$ solution:

\begin{equation}
\sum_{i=1}^N\frac{\left\{(x_i-\bar{x})^2 -
(\sigma_i^2+\sigma_Q^2)\right\}}{(\sigma_i^2+\sigma_Q^2)^2}=0
\end{equation}

In the case of zero measurement errors,  this reduces to the standard form:

\begin{equation}
\lim_{\sigma_i\rightarrow 0}\sigma_Q^2=\sum_{i=1}^N(x_i-\bar{x})^2/N
\end{equation}

In the case of identical measurement errors ($\sigma_m = constant$)
this reduces to the form used by Nandra et al (1997):

\begin{equation}
\lim_{\sigma_i\rightarrow
\sigma_m}\sigma_Q^2=\sum_{i=1}^N\left\{(x_i-\bar{x})^2 - \sigma_m^2\right\}/N
\end{equation}

In general however, Equation 7 can only be solved analytically for the
case where $N\leq2$.  For a larger number of bins we must find the
maximum likelihood estimate numerically.  Two contrasting examples of
light curves and their likelihood functions are given in Figures 1 and
2, from QSOs of high and low flux respectively. From these, maximum
likelihood estimates for $\sigma_Q$ can be evaluated with suitable
error bounds. The maximum likelihood estimate for $\sigma_Q$ is
obtained by locating the peak of the likelihood curve, while the error
bounds are obtained in the standard way by assuming that the
likelihood curve has a Gaussian shape. 

This method can either be applied to individual light curves or an
ensemble. In comparing values of $\sigma_Q$, however, it is important
to be aware of the effects of different total integration times and
sampling.  In addition, one might expect cosmological time dilation to
modify the results.  A detailed examination of these effects can be
found in Appendices A \& B. We find that one can apply a correction to
allow for these changes, but only under the assumption of a particular
underlying power spectrum. We therefore carried out our analysis with
and without these timescale corrections, but overall we found that
they have no significant effect on the results.

\begin{figure}
\caption{Light curves of the 12 QSOs showing evidence for variability
with $> 95$ per cent   significance.}
\centering
\centerline{\epsfxsize=7.2
truecm \figinsert{}{0.0pt}
}
\centerline{\epsfxsize=7.2
truecm \figinsert{}{0.0pt}
}
\centerline{\epsfxsize=7.2
truecm \figinsert{}{0.0pt}
}
\centerline{\epsfxsize=7.2
truecm \figinsert{}{0.0pt}
}
\centerline{\epsfxsize=7.2
truecm \figinsert{}{0.0pt}
}
\centerline{\epsfxsize=7.2
truecm \figinsert{}{0.0pt}
} 
\end{figure}

\begin{figure}
\caption{Showing the maximum likelihood estimates for the intrinsic
variability as a function of flux for the 86 QSOs. Timescale
corrections have not been applied. The point size is proportional to
the object flux, to aid in distinguishing objects of higher S/N in
Figures 5, 6, 8 and 9. } \centering \centerline{\epsfxsize=7.5 truecm
\figinsert{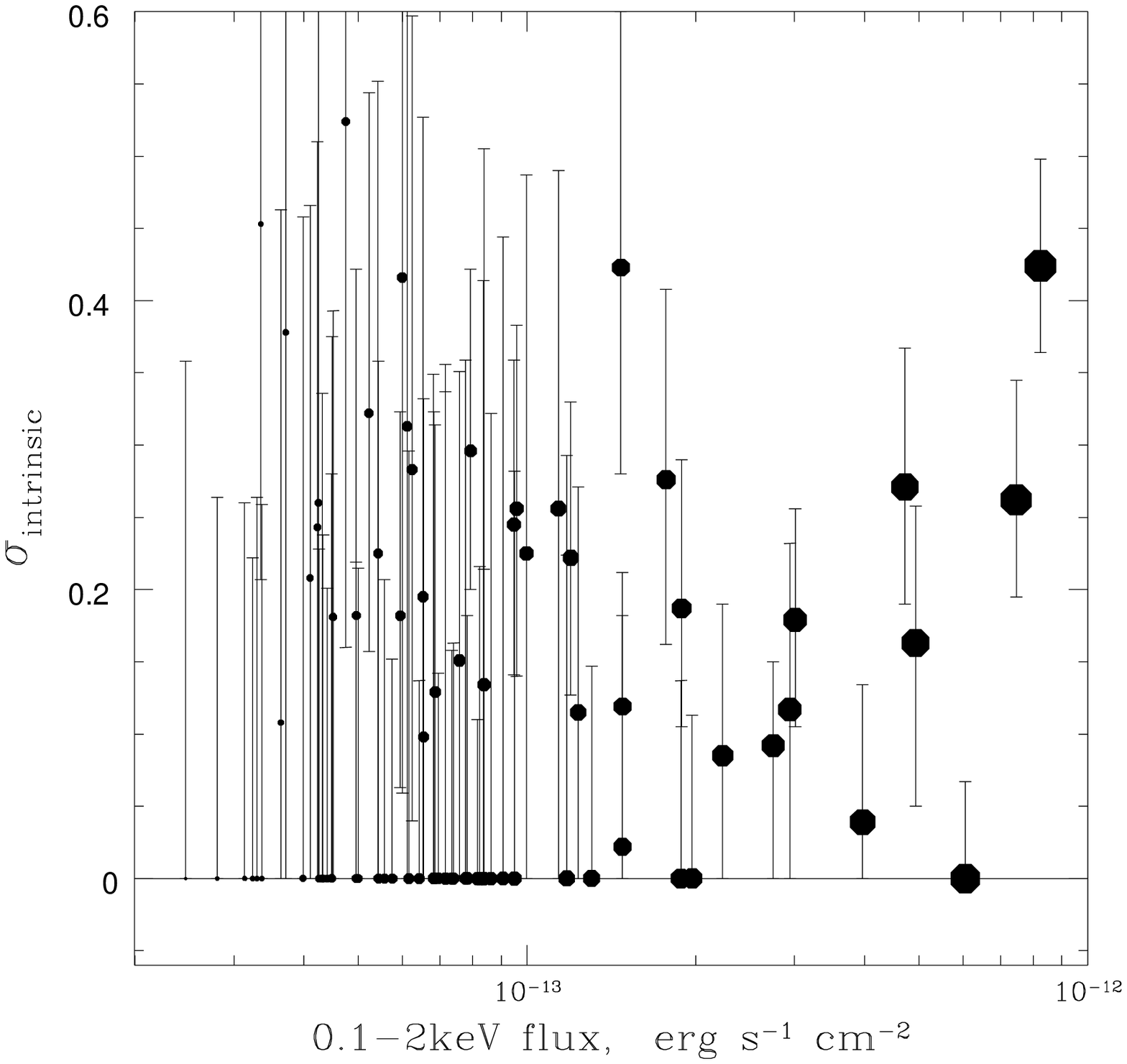}{0.0pt}}
\end{figure}

\section{The results}

\subsection{Detecting individually variable QSOs}

Before applying the maximum likelihood technique, a simple $ \chi^{2}$
test was performed on each light curve to test for variability against
the null hypothesis that the flux remains constant:

\begin{equation}      
\chi ^ 2 =\sum_{n}\frac{\left[I_{n}- \left\langle{I_n}\right\rangle
\right]^2}{\left\langle\sigma^2_{n}
\right\rangle}
\end{equation}        

The results show that only 12 of the 86 QSOs show individual evidence
for variability with a significance $> 95$ per cent, of which 4 are
significant at $> 99$ per cent.  The light curves for these QSOs are
displayed in Figure 3. As expected, the fraction of significantly
variable objects increases at higher flux. Of the 10 brightest QSOs
for example, 5 show variability at $> 95$ per cent significance, with
a further 2 showing variability at a significance of 90 per cent,
suggesting that a large fraction of all QSOs would show variability
given sufficient signal-to-noise.

It is important to emphasise that the non-detection of variability in
the remaining QSOs does not imply that no variability is occurring.
Any intrinsic variations which are present are simply overwhelmed by
photon noise. The detection of significant variability depends on the
amplitude of intrinsic fluctuations and on the signal to noise. Since
we can measure the photon noise, it should therefore be possible to at
least place an upper limit on the amplitude of intrinsic fluctuations
in every QSO.  Only then can we make quantitative statements about the
``typical'' variability in the population.

\begin{figure}
\caption{(a) Maximum likelihood estimates for the variability
amplitude as a function of luminosity for the 86 QSOs.  The point size
is proportional to the object flux (see Figure 4) in order to
highlight objects with higher S/N.  In (b) we display the results in
ensemble form, with unfilled squares showing the minor effect of not
applying timescale corrections.}  \centering
\centerline{\epsfxsize=7.5 truecm \figinsert{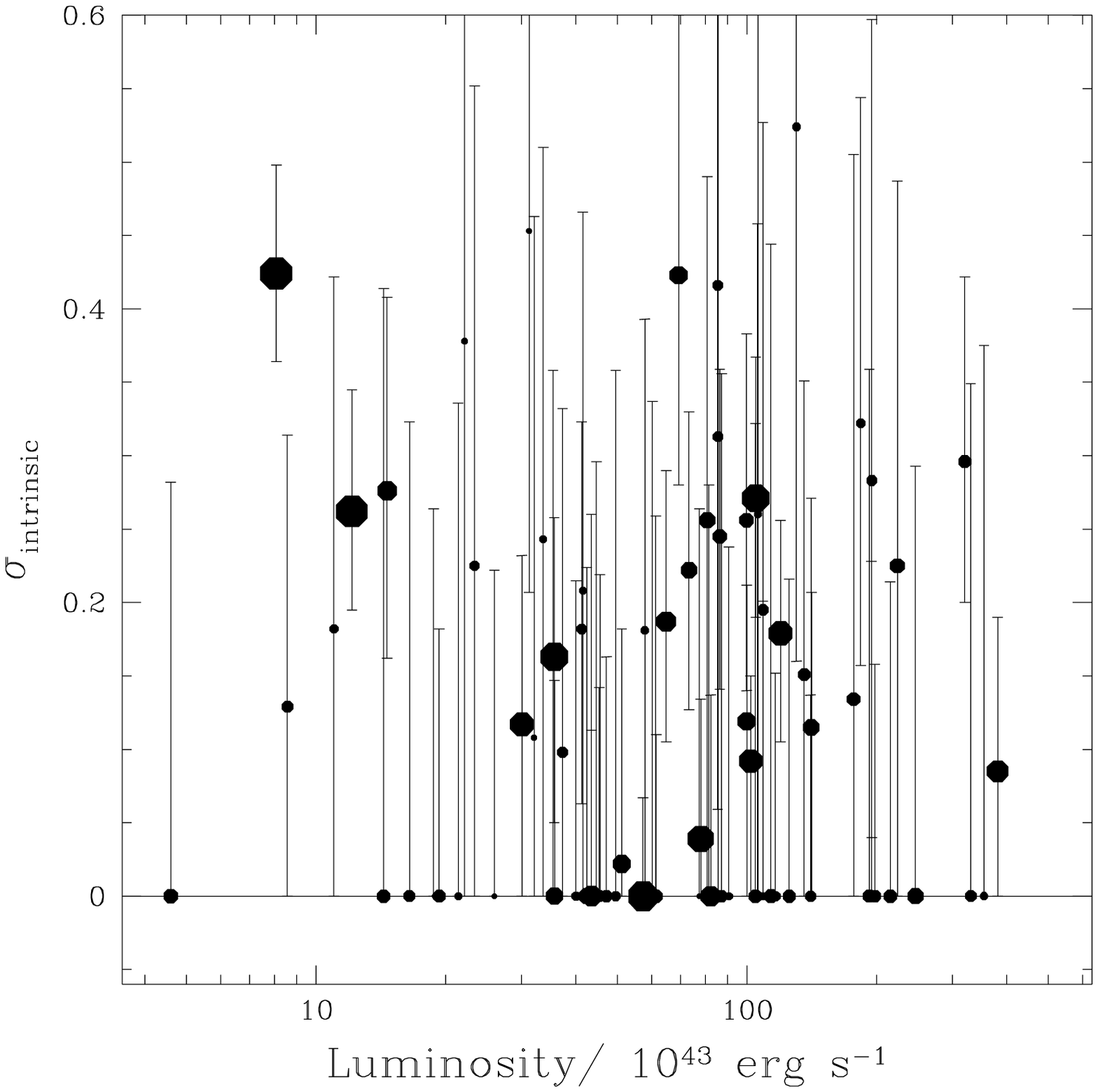}{0.0pt}}
\centerline{\epsfxsize=7.5 truecm \figinsert{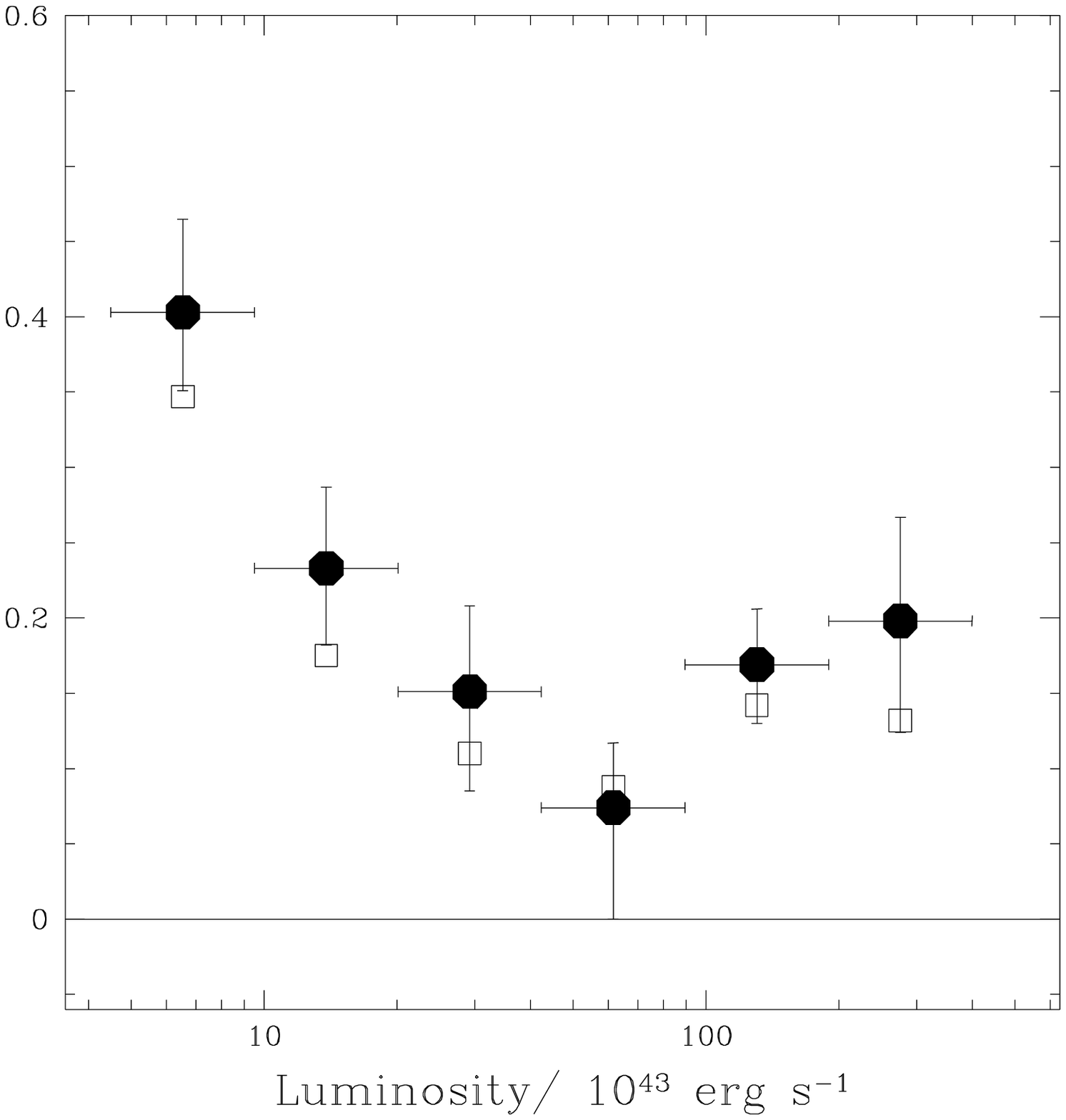}{0.0pt}}
\end{figure}

\begin{figure}
\caption{(a) Maximum likelihood estimates for the variability
amplitude as a function of redshift for the 86 QSOs.  The point size
is proportional to the object flux (see Figure 4) in order to
highlight objects with higher S/N.  In (b) we display the results in
ensemble form, with unfilled squares showing the minor effect of not
applying timescale corrections.}  

\centering
\centerline{\epsfxsize=7.5 truecm \figinsert{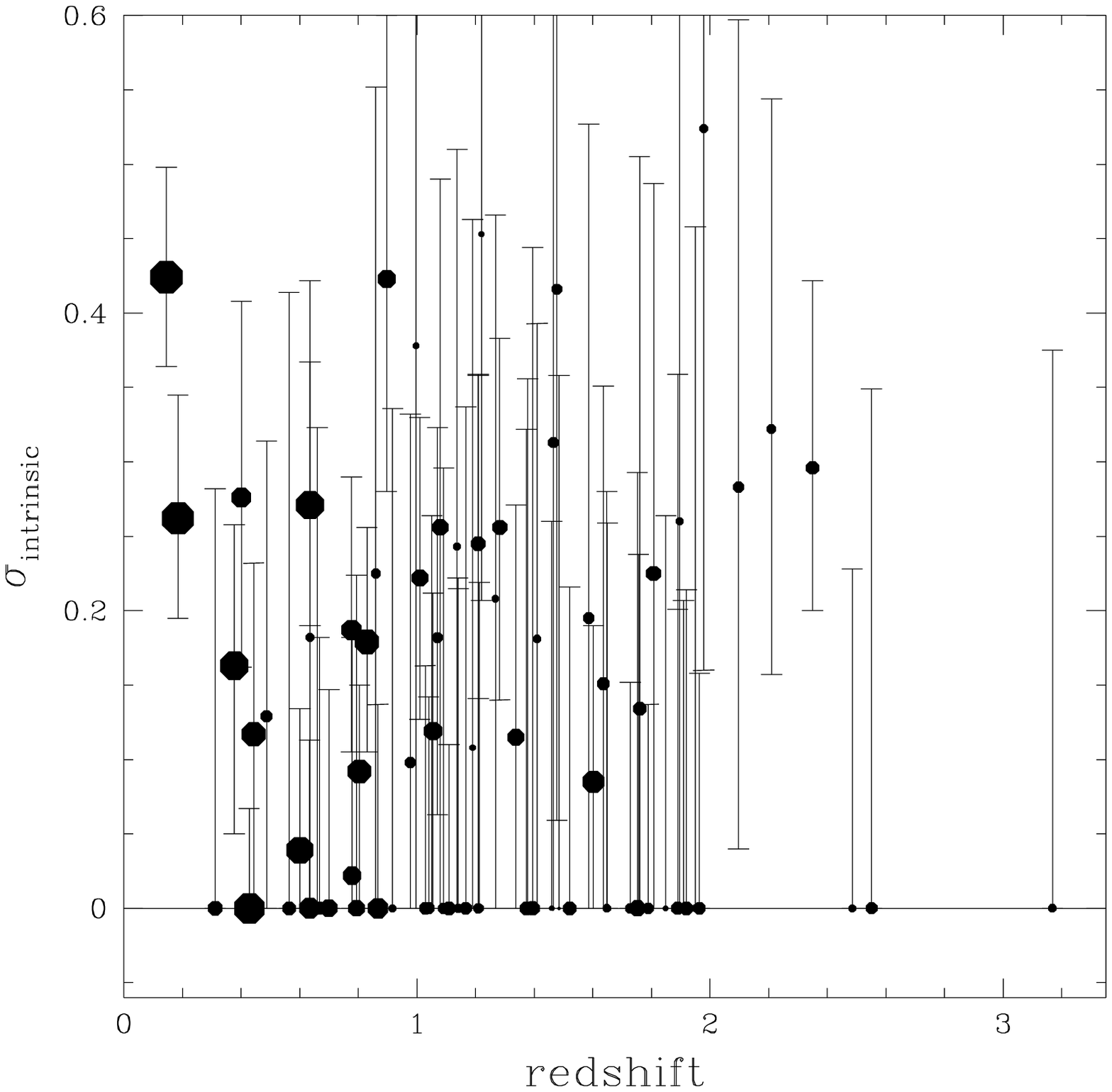}{0.0pt} }
\centerline{\epsfxsize=7.5 truecm \figinsert{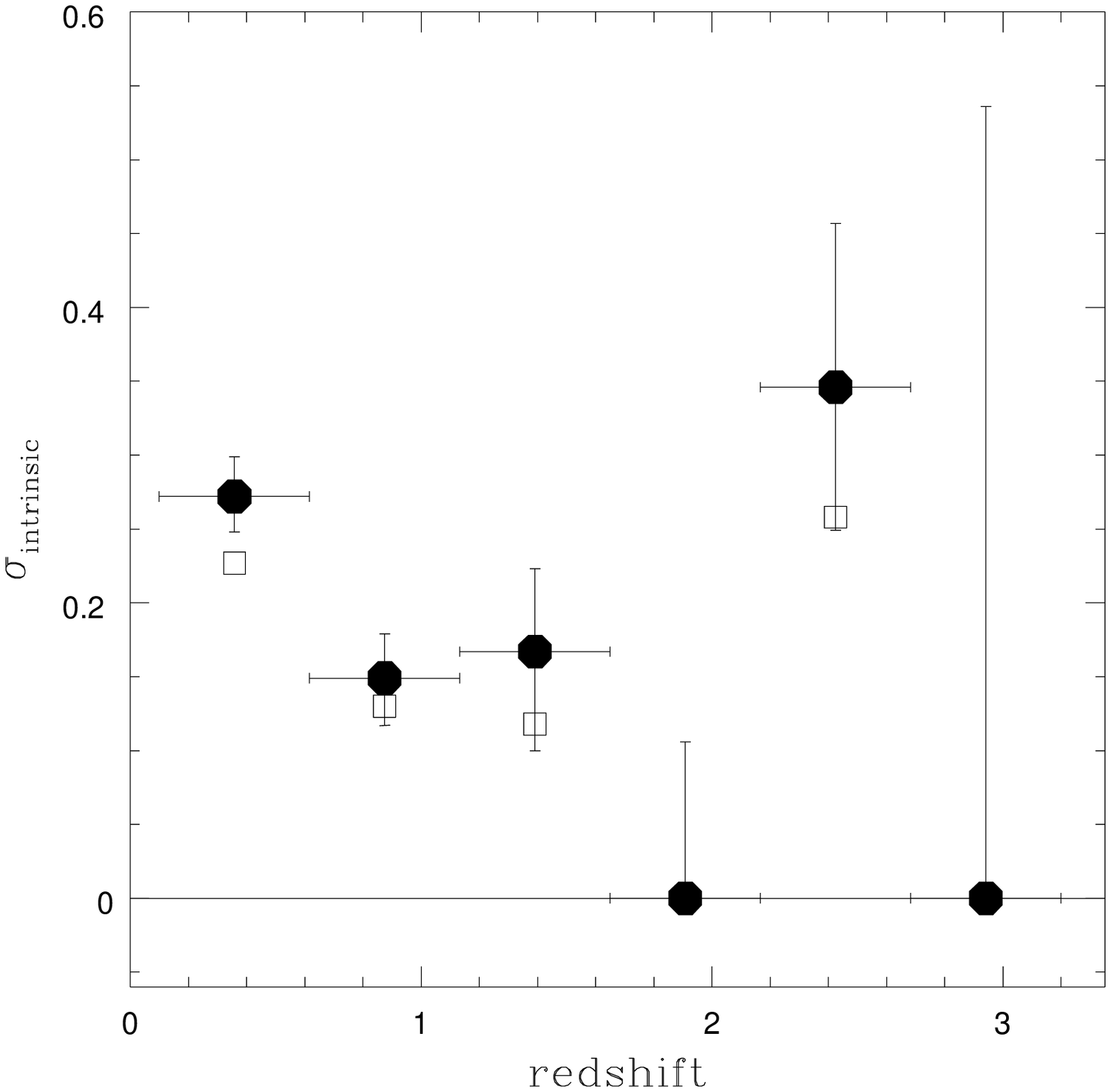}{0.0pt}}
\end{figure}

\subsection{The maximum likelihood results}

Numerically solving Equation 7 for each of the 86 QSOs, the maximum
likelihood estimates for the intrinsic variability amplitude are
displayed in Figure 4 (with $68$ per cent confidence regions),
plotting $\sigma$ as a function of flux. The timescale corrections
outlined in Appendix A are not applied to these raw data points, but
these have no major effect on the results (see below).  As expected,
for most faint QSOs individually we can only estimate upper limits for
the variability amplitude, but nevertheless the brighter QSOs and the
fainter objects combined would seem to suggest a typical variability
amplitude significantly above zero (with $\bar{\sigma} \sim 0.2$).

In Figures 5(a) \& 6(a) we display the variability amplitude as a
function of luminosity and redshift. We split the data into 6 equally
spaced luminosity and redshift bins and measure the ensemble variance,
applying the same maximum likelihood technique to the combined light
curves (Figure 5(b) and 6(b)).  The ensemble results are displayed
with and without the timescale corrections, which are dependent on the
assumed power spectrum. Although the net effect of these corrections
is to boost $\sigma$ upwards, they clearly have no significant effect
on the results.

We find a clear  detection of variability in the ensemble of
QSOs, with evidence for trends with luminosity and redshift. A
$\chi^2$ test was applied to test the null hypothesis that the scatter
in the 5 points on Figures 5(b) and 6(b) is consistent with their
error bounds.  For both the redshift and luminosity data, we can
reject the null hypothesis with $>99$ per cent confidence.

\subsection{Correlations With Redshift And Luminosity}

\begin{figure}
\centering \centerline{\epsfxsize=8truecm
\figinsert{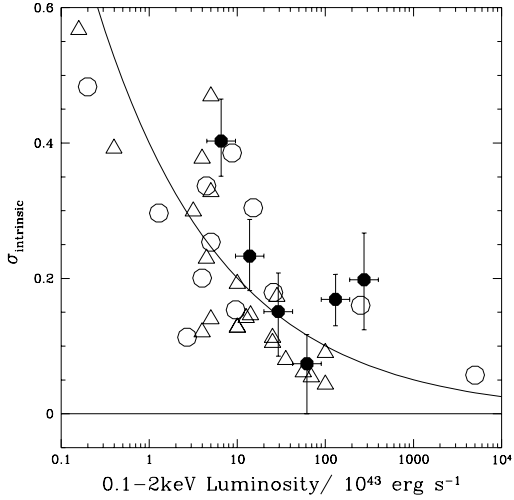}{0.0pt}}
\caption{A comparison of the {\em ROSAT} results with the findings of
Lawrence \& Papadakis (1993) (circles) and Nandra et al (1997)
(triangles).  The best fitting power-law of the form $\sigma^2\propto
L_x^{-0.6}$ is shown (not fitted to the {\rm ROSAT} QSOs).  To allow a crude
comparison, values of $\sigma$ have been normalised to timescales
below 1 week using a temporal power spectrum of the form $f^{-1.5}$
and luminosities have been converted to the $0.1-2\,$keV band.}
\end{figure}

Previous studies of very local ($z<0.1$) AGN found significant
evidence for a relationship variability amplitude and luminosity,
parameterised by the form $\sigma^2\propto L_x^{-{\rm(0.6\pm0.1)}}$
(Lawrence \& Papadakis 1993, Green \& McHardy 1993, Nandra et al
1997).  In Figure 7 we compare our ensemble QSO results with the work
of Papadakis \& Lawrence (1993) and Nandra et al (1997) (although we
note that there is considerable overlap between these local samples).
Similar results have recently been found by Turner et al (1999). To
allow a meaningful comparison we convert luminosities to our
$0.1-2$\,keV band (assuming a spectral shape of the form $\alpha=1.0$
above 1\,keV and $\alpha=1.5$ below 1\,keV).  Values of
$\sigma$ were corrected to timescales of 1 week (assuming a power
spectrum of the form $P(f)\propto f^{-1.5}$).  Given the uncertainties
in these conversions, and the fact that previous variability studies
were conducted at higher X-ray energies, we will be cautious in making
direct comparisons.  However, we note that the amplitudes found in our
QSOs are broadly consistent with these local AGN.  This would support
the hypothesis that the driving force in any variability trend is
actually the QSO luminosity, rather than the redshift. Formally,
however, a power-law gives a very poor fit to the  data (see Table 1).

An intriguing result is the apparent minimum at
$L_x\sim5\times10^{44}$erg s$^{-1}$, with an upturn towards higher
luminosities. No evidence for such an upturn has been seen among local
AGN.  To allow a fairer comparison, we split our sample into `local'
and `distant' QSOs, dividing at $z=0.5$ (Figures 8,9). We find that
while the local QSOs show a clear anti-correlation with luminosity,
consistent with previous studies, the QSOs at higher redshift show no
evidence for this trend whatsoever. To parameterise this difference,
we perform power-law fits to these distributions (see Table 1; similar
results are obtained if we fit to the binned, ensemble data).  Unlike
the full sample, we find that a power-law can describe these separate
distributions well ($\chi^2_{red} \sim 1$) and we obtain good fits
with $\sigma^2\propto L_x^{-{\rm(1.5\pm0.2)}}$ for the low redshift
QSOs but a very different form for the higher redshift objects, with
$\sigma^2\propto L_x^{+{\rm(0.6\pm0.3)}}$ (i.e.  a possible {\it
positive} correlation with luminosity). There is a strong degeneracy
between luminosity and redshift, however, and it would clearly be
desirable to obtain a larger sampling of the $L-z$ plane.  If one
restricts the comparison to only the 3 overlapping luminosity bins,
for example, a $\chi^2$ test rejects the hypothesis that the two
distributions are the same at only the $95$ per cent confidence level.
We therefore urge caution in over-interpreting these results, but we
will explore some of the possible consequences in Section 5.4.

\begin {table}
\begin {center}
\caption {Showing power-law fits to the distribution of $\sigma^2$ vs
luminosity for the {\rm ROSAT} QSOs.}
\begin {tabular}{||c|c|c||}
Data & Best fit & $\chi^2_{red}$ \\
\hline
All QSOs & $\sigma^2\propto L_x^{-{\rm(0.8\pm0.2)}}$ & 2.5 \\
$z < 0.5$ & $\sigma^2\propto L_x^{-{\rm(1.5\pm0.2)}}$ & 1.2  \\
$z > 0.5$ & $\sigma^2\propto L_x^{+{\rm(0.6\pm0.3)}}$ & 0.8 \\
\hline
\end{tabular}
\end{center}
\end{table}

\section{Constraining Physical Models}

Although we find some evidence for an upturn in the $\sigma/L_x$
relation at high luminosities, it is now well established that a
strong anti-correlation exists between these two parameters, at least
for local AGN. We will now explore some possible physical implications
of this trend, returning to the possible consequences of an upturn in
Section 5.4. 

The most plausible explanation for the observed anti-correlation is
that high luminosity sources have larger emitting regions.  It is
currently unclear whether this is due to a single, coherent but
erratically varying source or a large number of independent flaring
regions (Abramowicz et al 1991). The incoherence in the light curves
of $\em EXOSAT$ AGN would suggest a stochastic random process governed
by a large number of variables (Krolik, Done \& Madejski 1993), as
would the lack of any obvious timescale (Lawrence et al 1987).  We now
investigate how the observed relation $\sigma^2\propto
L_x^{-{\rm(0.6\pm0.1)}}$ can be used to constrain two classes of
models for AGN variability and in turn discuss the implications for
QSO evolutionary models.

\subsection{The single coherent oscillator}

In the case where the X-ray variability is caused by global, coherent
changes in the emitting region, it is reasonable to assume that the
luminosity of the source will be related to its physical size. The
geometry of the emitting region is unclear however, and could be
spherical, disk shaped etc., and hence we assume a general
relationship between the X-ray luminosity and the characteristic
radius, $R$:

\begin{equation}
L_x \propto R^{\beta}
\end{equation}

Over the range of frequencies we are interested in, we will assume a
temporal power spectrum of the form $P(f) \propto  f^{-\alpha}$ (where
$\alpha\simeq1.5$) as observed in local AGN (Lawrence \& Papadakis
1993).  Assuming self similar scaling, timescales will be scaled in
direct proportion to the size of the source, and hence in larger
sources the temporal power spectrum will simply be shifted in the
frequency direction. Self similar scaling requires conservation of the
total integrated fractional power, but the power measured over a
narrow frequency range will change.  By integrating the new power
spectrum over the observed frequency range one can relate the {\em
observed} variance to the size of the source:

\begin{eqnarray}
\sigma^2 & =         & \int_{f_1}^{f_2}P(f)df           \\
         & \propto    &  f_1^{1-\alpha} \hspace{1cm}  (f_1 \ll f_2, 
\alpha \neq 1) \\
         & \propto   &  R^{1-\alpha} \\
         & \propto   &  L_x^{(1-\alpha)/\beta} 
\end{eqnarray}

Thus one can readily reproduce the observed anti-correlation with
luminosity ($\sigma^2\propto L_x^{-{\rm(0.6\pm0.1)}}$) given the
measured power spectrum slope ($\alpha\simeq1.5$) if the geometry of
the source is such that $L_x\propto R^{ 0.7-1.0}$.  This would be
consistent with a ring of X-ray emitting material rather than a disk
or spherical cloud.  We note, however, that alternative geometries
could also reproduce the observations if we allow a radial decline in
the X-ray emissivity (e.g. free-free emission in a spherical cloud of
material if the electron density falls off at large radii). In
conclusion therefore, the observed anti-correlation can be reproduced
in this model and in principle could be used to constrain the geometry
of the emitting material.

\begin{figure}
\caption{(a) Variability amplitude as a function of luminosity for the
low redshift ($z<0.5$) QSOs. In (b) we display the results in ensemble
form.}  \centering \centerline{\epsfxsize=7.5 truecm
\figinsert{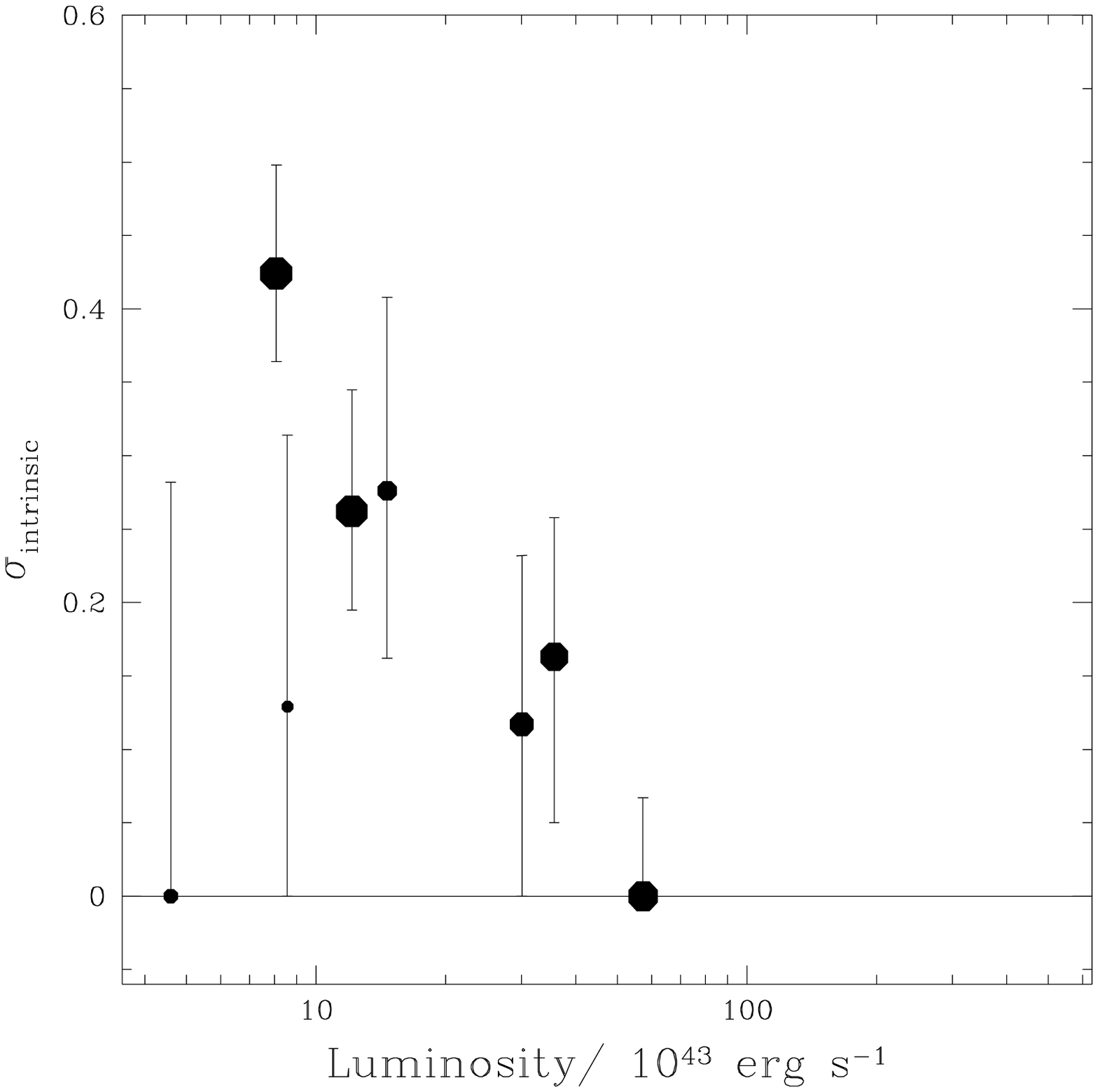}{0.0pt}} \centerline{\epsfxsize=7.5 truecm
\figinsert{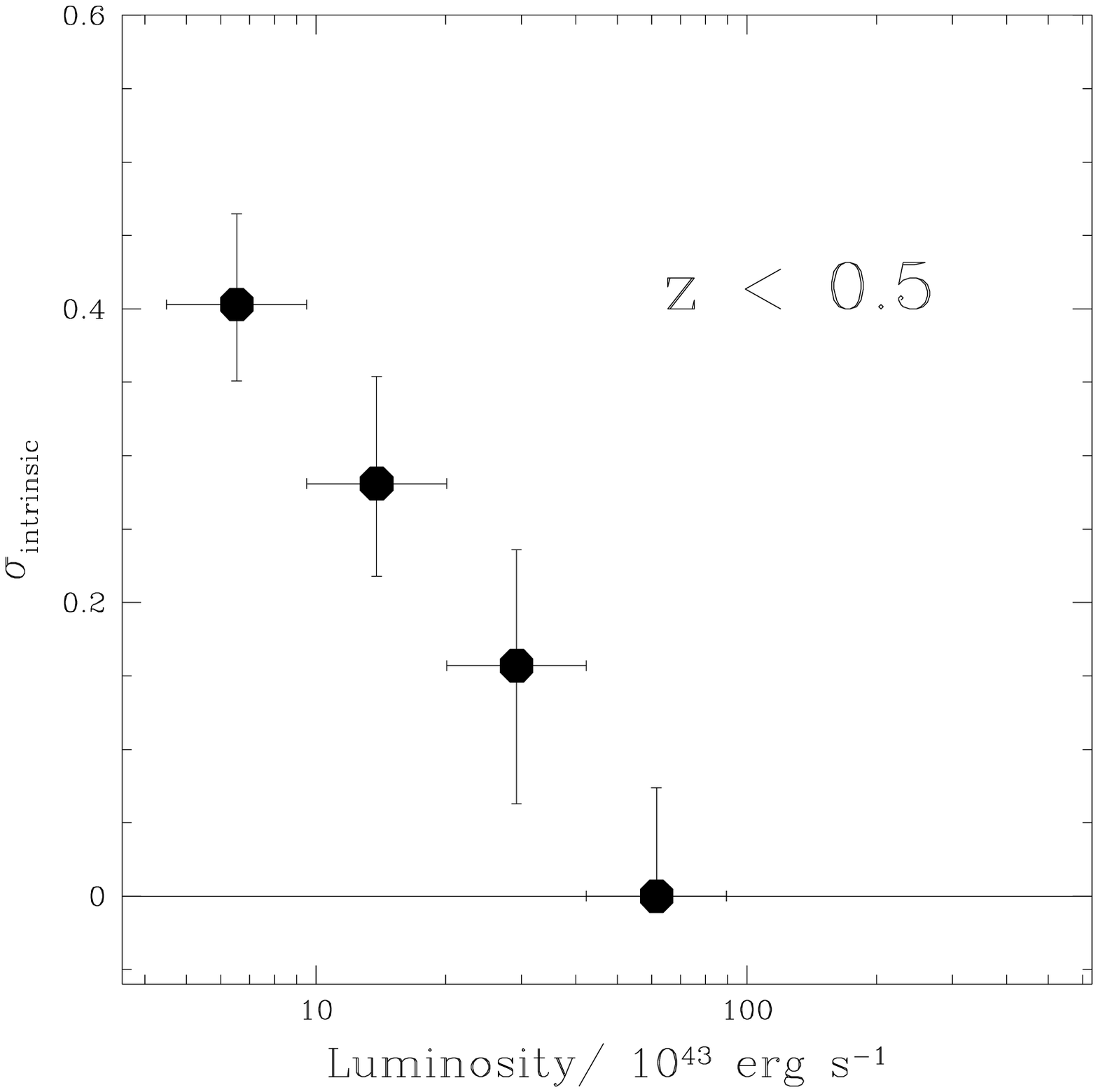}{0.0pt}}
\end{figure}

\begin{figure}
\caption{(a) Variability amplitude as a function of luminosity for the
high redshift ($z>0.5$) QSOs. In (b) we display the results in ensemble
form.}  \centering \centerline{\epsfxsize=7.5 truecm
\figinsert{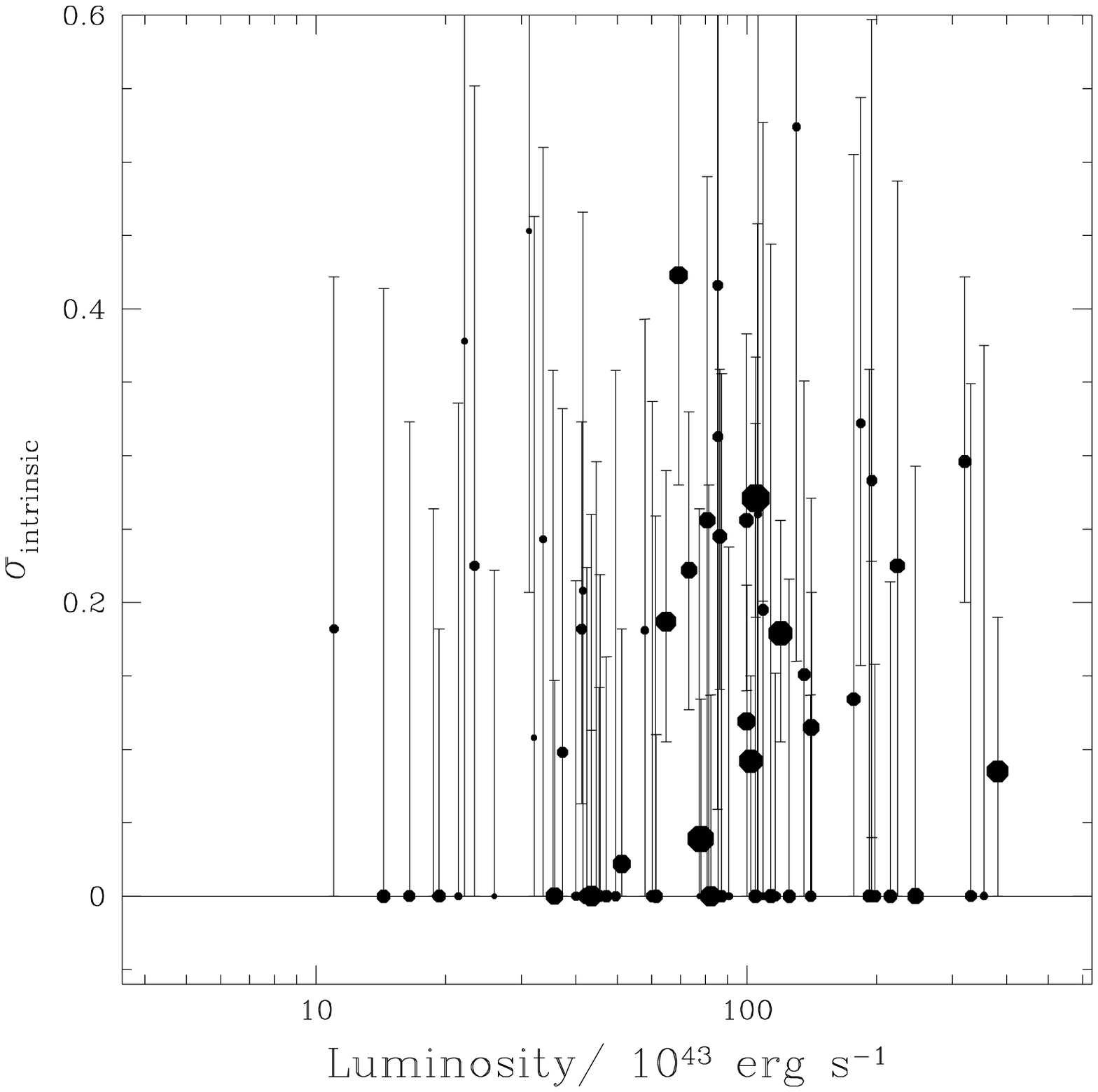}{0.0pt}} \centerline{\epsfxsize=7.5 truecm
\figinsert{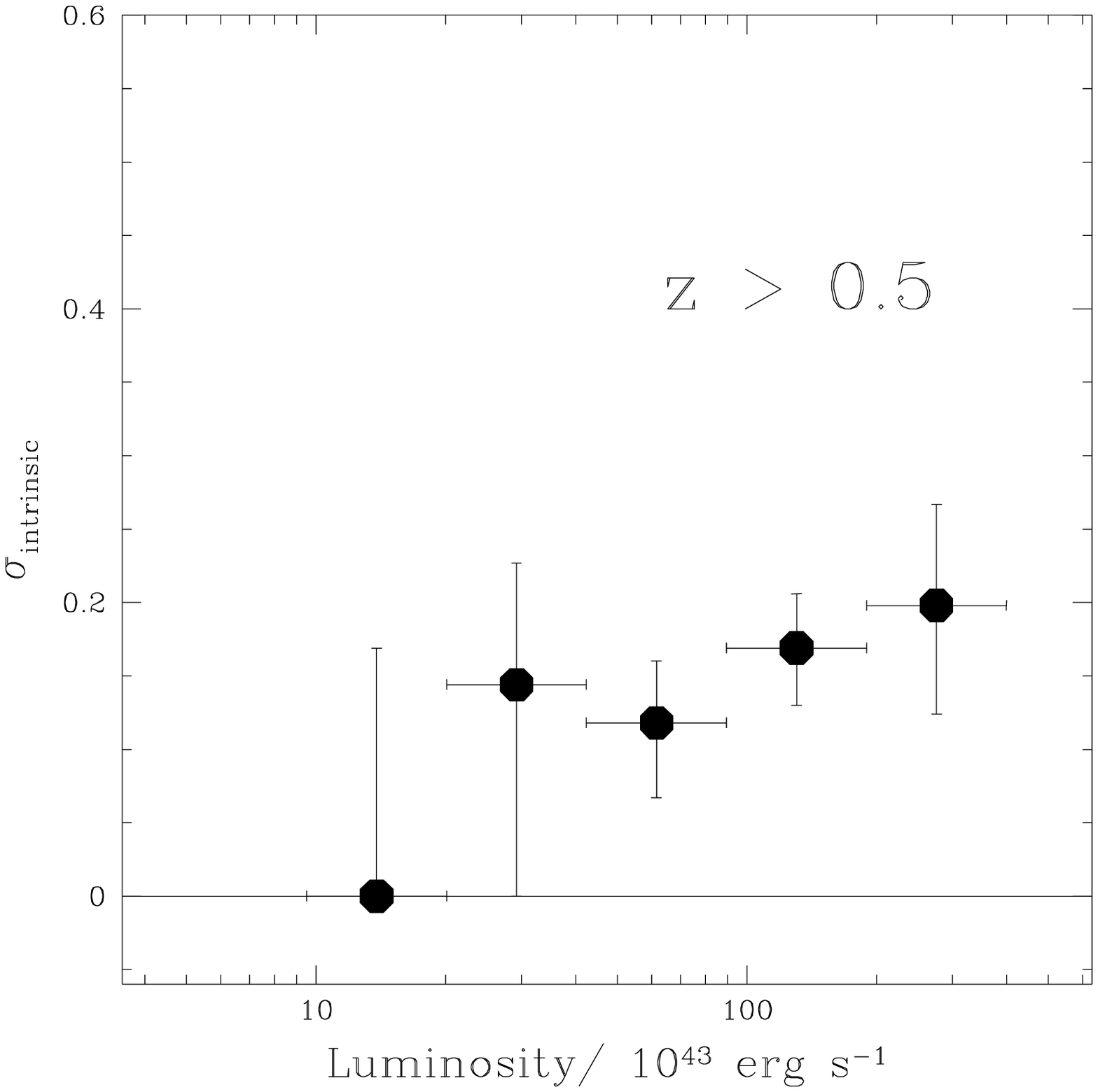}{0.0pt}}
\end{figure}

\begin{figure}
\centering
\centerline{\epsfxsize=9 truecm \figinsert{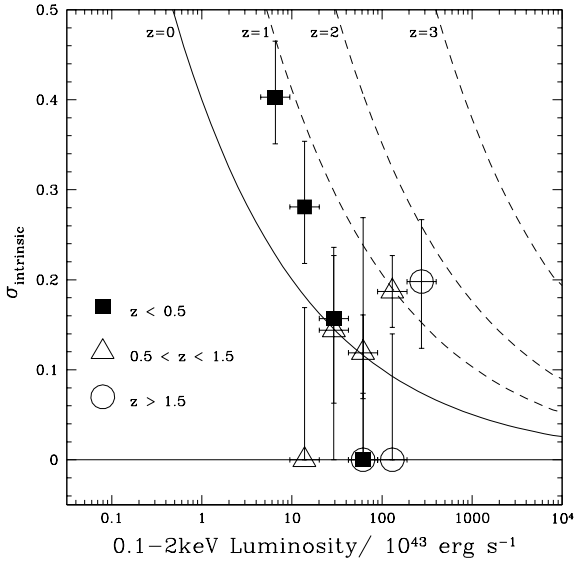}{0.0pt}}
\caption{Comparison with the predictions of a long-lived single
oscillator model. The local trend between variability amplitude and
source luminosity ($\sigma^2\propto L_x^{-0.6}$; solid line - see
Figure 8) is evolved to predict the relationship at high redshift
(dashed lines).  For comparison, the {\rm ROSAT} QSOs are superimposed,
split into 3 redshifts ranges.  }
\end{figure}

\subsection{The independent hot spot model}

In this class of model, the variability is caused by independent flaring
events. More luminous sources will simply have more flaring regions
and hence the r.m.s. fluctuations will be smaller.  If the
time-averaged luminosity from each flare is $\bar{l}$, the typical
number of flares at a given time will be related to the total 
X-ray luminosity, $L_x$ by:

\begin{equation}
N=L_x/\bar{l}
\end{equation}

Thus the fractional r.m.s. variation on the number of flares at a given time 
will be:

\begin{equation}
\sigma = \sqrt{N}/N \propto \left(\frac{L_x}{\bar{l}}\right)^{-0.5}
\end{equation}

If the flares are the same size in all QSOs, this would lead to
$\sigma^2\propto L_x^{-1}$, which is somewhat steeper than the
observed correlation. This could be overcome if higher luminosity
objects had more luminous flares, such that $\bar{l}\propto
L_x^{0.5}$. Hence this  model can also explain the observed anti-correlation.

\subsection{Constraining models of QSO evolution}

As outlined in Boyle et al (1987) there are at least two possible
models which can explain the strong evolution in the QSO
luminosity function:

\newcounter{count}
\begin{list}
{(\roman{count})}{\usecounter{count}}
\item The evolution of successive generations of short lived 
$(\sim 10^8 {\rm yr})$ QSOs in which the mean intensity of QSO cycles
diminishes. In this scenario most galaxies have been through a quasar
epoch at least once.
\item The evolution of individual, long-lived QSOs which uniformly dim over
time. In this scenario only $\sim 1\%$ of galaxies have ever undergone
quasar activity, but would develop very massive black holes.
\end{list}

In the first model, we expect a typical QSO to harbour a black hole of
mass $\sim 1\times10^{8} M_{\odot}$ regardless of redshift. Black
holes of this size allow the theoretically favoured accretion at
Eddington limit (Rees 1984) for a QSO at peak activity. Thus the size
of a typical emitting region is expected to be very similar at all
redshifts.

In the more controversial long-lived model, QSOs have poured out a
vast amount of energy over most of the age of the universe and should
therefore develop extremely massive black holes by the associated
accretion. This can be calculated by integrating the total energy
emitted by the QSO from it's formation until the present day (full
details given in Appendix C). If we assume an accretion efficiency
$\varepsilon=0.1$ after Rees (1984), an initial mass given by the
Eddington limit and a starting redshift of $z=3$ then we predict a
factor $\sim 20h_{50}^{-1}$ increase in black hole mass from $z=3$ to
$z=2$, with a further growth by a factor of $\sim 3h_{50}^{-1}$ from
$z=2$ to the present day (Figure C1).

In principle therefore, since variability traces the size scales in
the central regions of an AGN, one might be able to distinguish
between these models. The consequences depend on the variability
model:

{\bf Single Coherent Oscillator}: In this case, the amplitude of
variability is directly related to the physical size of the source
(Equation 14). The X-ray emission from AGN is generally thought to
arise a few Schwarzschild radii from the central black hole, and in
this metric the central mass is directly proportional to the size
scale.  Hence one can relate the amplitude of variability to the
central black hole mass, $M_{BH}$:

\begin{equation}
\sigma^2  \propto  R^{1-\alpha} 
\propto  M_{BH}^{-{\rm0.5}}           
\end{equation}

In the long-lived model this would lead to dramatic changes in the
variability properties of QSOs with redshift (at a fixed
luminosity). First, as the black hole mass of an individual QSO
decreases towards higher redshift, the variability will increase as
$\sigma\propto {M_{BH}}^{0.25}$ (most dramatically at $z>2$).  In
addition, because of the evolution in the luminosity function (in this
model due to the fading of individual QSOs) one should compare objects
which have followed a similar evolutionary path (i.e. the same
starting luminosity).  The observed evolution in the QSO luminosity
function (Boyle et al. 1994) would imply a decline in luminosity by a
factor of $\sim 30$ from $z=2$ until the present day.  Combining both
of these effects, the predicted changes are striking (see Figure 10).

The model predictions are, of course, strongly dependent on the
assumed accretion efficiencies, redshift of formation etc., but the
generic prediction is a rise in the mean variability (at fixed
luminosity) towards higher redshift.  Our data do not appear grossly
inconsistent with these trends, particularly at the highest
luminosities where the high redshift QSOs do indeed show evidence for
an upturn (see also Figures 8 \& 9).  At present, therefore, we
cannot, conclusively rule out a long-lived QSO model on the basis of
variability evolution. Further measurements are clearly required. In
particular, examining variability properties at high redshifts ($z
\sim 3$) would provide a conclusive test, where the differences are
most striking. This will be possible with Chandra and XMM.

In the alternative model, most galaxies have been through a quasar
phase (for perhaps $10^8$ years) and we expect a similar range of
black hole masses at any given epoch. This model is also consistent
with the current data, although we note that this would not explain
the apparent upturn in the $\sigma/L_x$ relationship for higher
redshift AGN.

{\bf Multiple hotspot model:} In this model, the variability amplitude
is determined by the number of independent flares.  Hence predictions
relating to evolutionary models will depend on the physics of the
flaring mechanism. If the number of flares is strictly governed by
only the source luminosity, then we expect no difference between the
long and short lived evolution models. However is seems plausible that
a physically larger emitting region will give rise to a greater number
of disconnected, independent emitting regions. Hence the similarity in
the variability trends at high and low redshift may also be sufficient
to rule out a long-lived evolutionary model.

\subsection{Explaining the high luminosity upturn}

In Section 4 we found evidence that the local anti-correlation between
variability and luminosity does not hold at the highest
luminosities. Although further observations are clearly required to
test the reality of this feature, we will nevertheless speculate on
some possible explanations.

Our results tentatively suggest that this feature is caused by
redshift evolution, since while our low redshift QSOs show the same
anti-correlation between $\sigma$ and $L_x$ as previous studies, the
higher redshift sources show very different behaviour. A simple
interpretation is that earlier QSOs (at a given luminosity) were
powered by less massive black holes compared to local AGN (or
alternatively were accreting more efficiently).

On the other hand, if luminosity is the driving parameter (rather than
redshift evolution) it could reflect an upper limit to the maximum
black hole size. Assuming accretion at close to the Eddington limit,
the upturn X-ray luminosity ($5\times10^{44}$ erg s$^{-1}$)
corresponds to a mass of $\sim 5\times10^7 M_{\odot}$. This is
strikingly similar to the mass at which the space density of active
black holes turns over sharply (e.g. see Padovani, Burg \& Edelson
1990).  This could, in principle, lead to a levelling off in the
$\sigma$ vs. $L_x$ relationship if we assume (not unreasonably) that
there is some scatter in the luminosity for a given black hole size.

\section{Variability of X-ray luminous galaxies}

In recent years, is has become clear from deep X-ray surveys that a
population of X-ray luminous narrow emission-line galaxies (NELGs)
might provide an explanation for the origin of the cosmic X-ray
background (Boyle et al. 1995, Roche et al. 1995, Carballo et
al. 1995, Almaini et al 1997, McHardy et al 1998). These galaxies are
typically $100$ times brighter than normal field galaxies, and appear
to have harder X-ray spectra than QSOs (Almaini et al 1996). The
nature of this X-ray emission is still controversial, but the presence
of massive starburst activity and/or obscured AGN has been postulated.

Here we investigate the X-ray variability of a small number of these
galaxies from our Deep $\em ROSAT$ Survey. Unfortunately most of these
galaxies are too faint to enable a meaningful variability
analysis. Only 6 of the 23 X-ray luminous galaxies presented in
Almaini et al (1996) meet the selection criteria outlined in Section
2. A $\chi^2$ test was performed on these light curves. The null
hypothesis of constant flux can only be rejected in the galaxy with
highest flux, namely GSGP4X:091, with a significance of $>95$ per
cent.  The light curve is displayed in Figure 11. Such variability
clearly favours an AGN origin for the X-ray emission. Applying the
maximum likelihood technique, we obtain variability amplitudes
$\sigma=0.31\pm_{0.15}^{0.19}$, consistent with the typical
fluctuations seen in low luminosity AGN (see Figure 5).  Combining the
light curves of the 5 fainter galaxies, we do not detect a variability
amplitude significantly above zero. Formally we obtain a $68$ per cent
upper limit of $\sigma <0.19$, from which we certainly cannot rule out the
presence of AGN-like variability.

\begin{figure}
\centering \centerline{\epsfxsize=5truecm
\figinsert{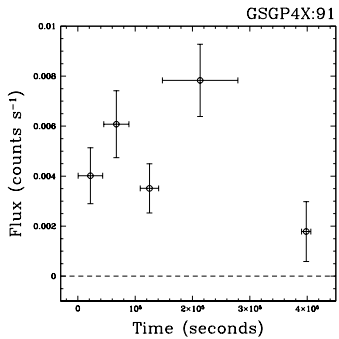}{0.0pt}}
\caption{Light curve of the  variable NELG GSGP4X:091}
\end{figure}

\section{Conclusions}

We have developed a technique for constraining the amplitude of
variability in light curves of low signal to noise. By treating each
(normalised) light curve as a `snapshot' from an ensemble, one
can obtain a maximum likelihood estimate for the typical variability
in the population.

We apply this technique to the light curves of 86 QSOs obtained as
part of the Deep $\em ROSAT$ Survey.  This represents the first large
study of X-ray variability in QSOs over a range in redshift.  On
timescales of several days, the amplitudes of variability are very
similar to those found in very local AGN.  We find evidence for trends
in the ensemble variability amplitude with both redshift and
luminosity. The trend with luminosity is broadly consistent with the
anti-correlation seen in local AGN, suggesting that this is the driving
parameter rather than redshift. 

Unlike local AGN, however, we find evidence for an upturn in the
variability at the highest luminosities ($L_x > 5\times10^{44}$erg
s$^{-1})$. We speculate on possible explanations for this upturn. If
luminosity is the driving parameter, it could reflect the turn over in
the mass spectrum of active black holes. This could also be caused by
redshift evolution, e.g. if luminous, high redshift QSOs contain
smaller black holes or accrete more efficiently. 

Finally, we find evidence for X-ray variability in an object
classified as a narrow emission-line galaxy, suggesting the presence
of AGN activity.

\section*{ACKNOWLEDGMENTS}

We are indebted to Alan Heavens for considerable help with statistical
matters, and to Rick Edelson, Niel Brandt and Andy Fabian for useful
discussions. OA would also like to thank Seb Oliver, Chris Willott and
Dave Alexander for an illuminating (if somewhat drunken) conversation
in Bologna.

\appendix

\section{Correcting for observation length, binning and time dilation}

Detailed Fourier analyses of local AGN have revealed
larger amplitude fluctuations at low frequencies, parameterised
by  power spectra of the form:

\begin{equation}
P(f) \propto f^{-\alpha}
\end{equation}

where $\alpha\simeq1.5$. Thus the amplitude of variability will depend
on the exact range of timescales that we sample. If we sample
frequencies in the range $f_{1} < f < f_{2}$, the intrinsic
variance will relate to the power spectrum as follows $(\alpha\neq1)$:

\begin{eqnarray}
\sigma^2 & = &\int_{f_{2}}^{f_{2}}P(f)df  \\
& \propto & f_{1}^{1-\alpha}-f_{2}^{1-\alpha}  
\end{eqnarray}

One can use this to correct for different sampling by normalising all
QSOs to the same frequency range.  We adopt a low frequency bound
$f_{1}$ corresponding to a timescale of 1 week ($1.65 \times
10^{-6}$ Hz). This reflects the typical observation length in our
QSOs.  We arbitrarily choose $f_{2}=10^{-3}$ Hz, corresponding to the
length of the shortest bins ($\sim 1000s$).

In order to normalise to these frequencies, 3 distinct corrections
must be applied to the amplitudes ($\sigma$) derived in Section 3.
The largest of these arises because the 7 fields are of differing
observation lengths, varying from $\sim 2$ days for the QSF3 field to
a maximum of $\sim 14$ days for the F864 field. We expect larger
amplitude variations for longer integration times. For a total
observation length $T$, assuming perfect sampling and $f_{1}\ll
f_{2}$, Equation A2 leads to:

\begin{equation}
\left\langle \sigma \right\rangle \propto T^{(1-\alpha)/2}
\end{equation}

A second bias arises because of the irregular gaps in the data and
from the re-binning described in Section 2. In particular, the bins in
many fainter QSOs are merged, which will reduce the observed variance
by averaging out high frequency variability.  Quantifying these
effects would be exceptionally difficult analytically because of the
irregular sampling. We therefore carry out simulations for each QSO in
order to quantify this behaviour. This is described in detail in
Appendix B. For the faintest QSOs with only 2 widely separated bins,
we find that these effects produce an average reduction in $\sigma_Q$
by a factor $\beta=1.34$ compared to a perfectly sampled light curve.

Finally a correction may also be applied for the effect of cosmological time
dilation. If we sample frequencies in a light curve in the range
$f_{1} < f < f_{2}$ then the intrinsic standard deviation in the
light curve will be given by Equation A3. If instead we observe the
redshifted frequencies $f'=f/(1+z)$ then the observed $\sigma$ will be
reduced by approximately $(1+z)^{-0.25}$. Thus the amplitude we have
measured for QSOs of redshift $z=2$ should be increased by a factor
$\sim 1.3$ for comparison with QSOs at zero redshift.

In principle we can correct for each  of these effects in order to
compare all QSOs within the same frequency range.  We therefore
normalise all values of $\sigma_Q$ as follows:

\begin{equation}
\sigma_{Q}^{norm} =(1+z)^{0.25}\times\beta \times (T_i/week)^{-0.25}
\times \sigma_{Q}^{data}
\end{equation}

When combining an ensemble of QSOs, this correction can be included in
Equation 7 (simply replace $\sigma_Q$ with $\sigma_Q^{data}$ as given
above, and solve for $\sigma_{Q}^{norm}$ with individual corrections
for each QSO).  Since these three corrections rely on an assumed form
for the power spectrum, which may not be entirely valid for the
general QSO population, we  also display
results based on the raw, uncorrected data for comparison.

\section{Simulating light curves}

A time process $x(t)$ may be represented by its Fourier integral:

\begin{equation}
x(t)=\int_{-\infty}^{+\infty}X(f)e^{2{\pi}ift}df
\end{equation}

The power spectrum of $x(t)$ is defined as the modulus squared of the
inverse Fourier transform (neglecting factors of $2\pi$):

\begin{equation}
P(f) = \left|\int_{-\infty}^{+\infty}x(t)e^{-2{\pi}ift}dt\right|^2
  = \left|X(f)\right|^2 
\end{equation}

Rather than assuming random variability (ie. white noise) for the
QSOs, we assume that all QSOs show the same power spectra as local
Seyfert galaxies, ie. $P(f) \propto f^{-1.5}$ (McHardy \& Czerny 1987,
Lawrence \& Papadakis 1993).  A general theoretical light curve can
then be constructed to satisfy this spectral form:

\begin{equation}
x(t)=\int_{-\infty}^{+\infty}\left|P(f)\right|^{1/2} e^{i[2{\pi}ft-\phi(f)]}df
\end{equation}

where the real function $\phi(f)$ represents the phase information
lost by the power spectrum method.  In a stochastic random process
governed by a large number of variables we expect no relation between
the phase and frequency of Fourier components, and indeed this has
been shown for a number of $\em EXOSAT$ AGN (Krolik, Done \& Madejski
1993).  The light curves were therefore simulated using a random phase
$\phi(f)$.  Sampling $x(t)$ over discrete time intervals $t_k$, the
Fourier integral collapses into the discrete Fourier transform and we
obtain:

\begin{equation}
x(t_k)\propto\sum_{f_{1}}^{f_{2}}
\left|P(f)\right|^{1/2}\cos\left[2{\pi}ft_k-\phi(f)\right]
\end{equation}

The lowest contributing frequencies are chosen at $10^{-6}$ Hz.  The
high frequency cut off (the Nyquist frequency) was chosen at an
arbitrarily high value of $10^{-2}$ Hz, close to the limit of power
spectrum observations (Nandra et al 1997). To obtain the required
amplitude of variability, the light curves are scaled and a flat
d.c. component added to give a total fractional variance of the
required amplitude.  The resulting light curve is then corrected to
allow for the redshift of the QSO (viz. all times multiplied by $1+z$)
and a random section is sampled and rebinned in exactly the same way
as the real data. 1000 such simulations are carried out for each QSO
in order to determine the correction factor, $\beta$, which gives the
mean change in the intrinsic standard deviation $\sigma$ as a result
of the binning structure:

\begin{equation}
\beta=\sigma_{intrinsic}/\left\langle\sigma_{binned}\right\rangle
\end{equation}

For the 86 QSOs in our sample, values of this correction factor lie
in the range $1<\beta<1.34$ with the largest values for the faintest
QSOs with only 2 widely spaced
temporal bins.

\section{The evolution of black hole mass in a long-lived QSO model}

Averaged over an accretion timescale, the luminosity of a QSO
relates to its accretion rate via:

\begin{equation}
L(z)=\epsilon\dot{M}c^2 = 5.7\times10^{46}\epsilon\left(\frac{\dot{M}}
{M_{\odot}yr^{-1}}\right)
\end{equation}

where $\epsilon$ is the accretion efficiency, taken to be $\sim 10\% $
after Rees (1984), and L is the bolometric luminosity emitted across
the entire electromagnetic spectrum.  For a typical radio quiet quasar
the bolometric correction from $0.1-2$\,keV is approximately a factor
of 12 from gamma-ray to far infra-red wavelengths.

In the long-lived QSO model, the evolution of the QSO luminosity function
(e.g. Boyle et al 1994) directly describes the evolution of individual QSOs:

\begin{equation}
L\propto(1+z)^3  \hspace{1.2cm} z<2
\end{equation}

At higher redshifts, studies of optical (and radio) quasars suggest
that the luminosity function is roughly flat during the epoch $2<z<3$,
with a rapid decline beyond redshift $z=3$ (Schmidt, Schneider \& Gunn
1995). For the purposes of a long-lived model, we therefore assume
QSOs began accreting at $z=3$ and estimate the initial mass of the
black hole by assuming that the system began by radiating at the
Eddington luminosity (Rees 1984).

\begin{equation}
L_{max} = L_{Edd}\sim1.3\times10^{38}\frac{
M(z=3)}{M_{\odot}}{\rm erg}s^{-1}
\end{equation}

In order to relate the evolution in time with the evolution
in redshift, we use the standard relation from a Robertson-Walker metric
($\Lambda=0, q_o=1/2$):

\begin{equation}
t=2/3H_o^{-1}(1+z)^{-3/2}
\end{equation}

From the above, one can obtain the evolution of mass with redshift:

\begin{eqnarray}
\frac{d M}{d z} & = & \frac{d M}{d t}.\frac{dt}{dz} \\
   & = & \frac{-H_o L(z)}{\epsilon c^2}(1+z)^{-5/2} \\
   & = & \frac{-H_o L_{max}}{27\epsilon c^2}(1+z)^{1/2} 
\end{eqnarray}

We can integrate this relation to obtain the mass as a function of
$z$, where we fix the constant of integration using Equation
C3. This is shown in Figure C1. For the $z<2$ epoch, this evolution is
parameterised as follows:

\begin{equation}
M(z)  = \frac{L_{max}}{10^{38}}
 \left[13+0.7\epsilon^{-1}h_{50}^{-1}
(5.2-(1+z)^{3/2})\right]   
\end{equation}

\begin{figure}
\centering
\centerline{\epsfxsize=7.5 truecm \figinsert{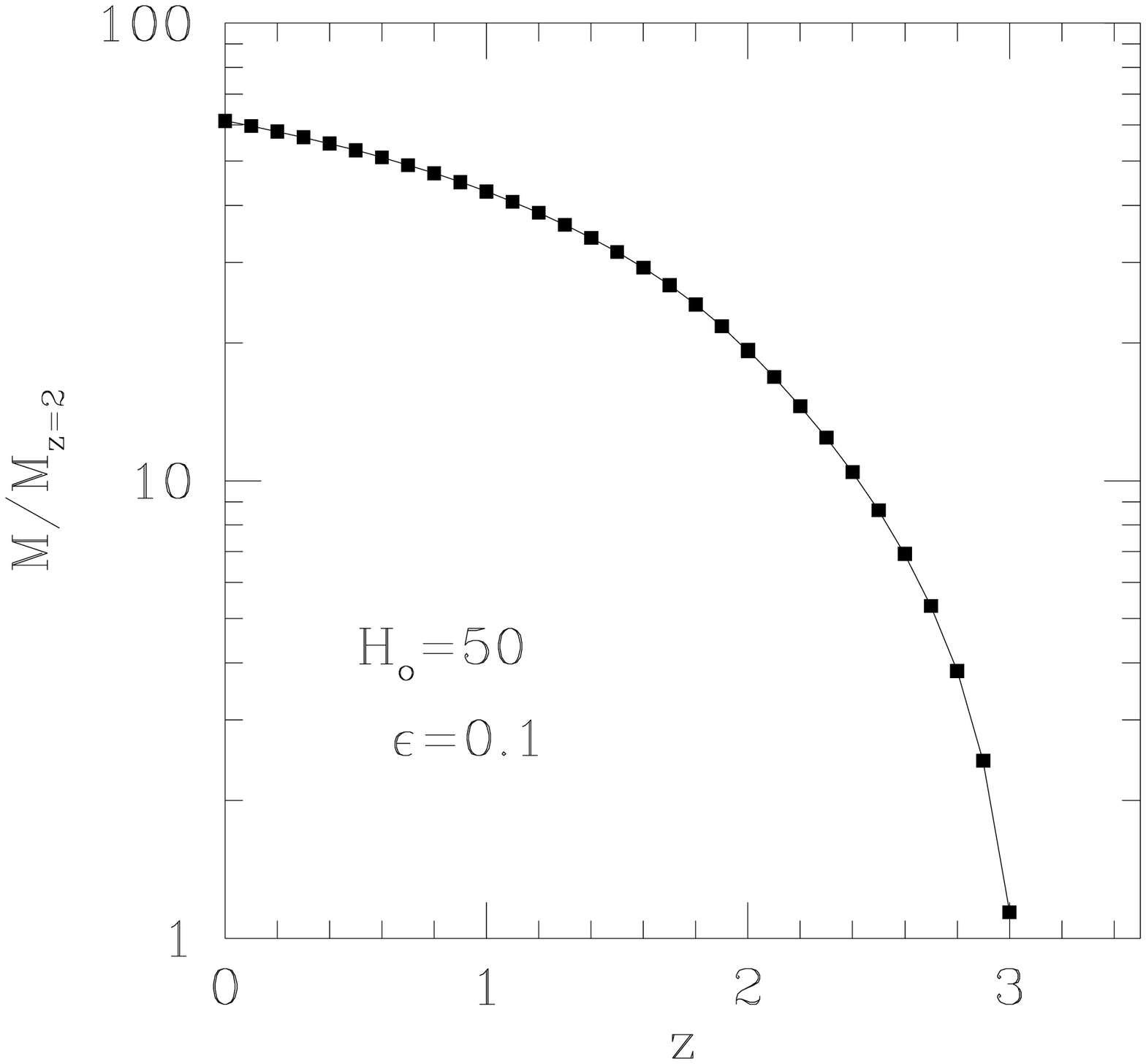}{0.0pt}}
\caption{Showing the predicted growth in black hole mass from $z=3$
until the present day, assuming a long-lived QSO evolution model, with
an initial mass given by the Eddington limit.  }
\end{figure}

\end{document}